\documentclass[floats,floatfix,showpacs,amssymb,prd,twocolumn,superscriptaddress,nofootinbib,nolongbibliography,reprint]{revtex4-2}

\usepackage{amssymb,amsmath,verbatim,mathtools,needspace,enumitem,etoolbox,graphicx,physics,microtype,afterpage,bigints,gensymb,tabularx,xspace}

\usepackage[dvipsnames, usenames]{xcolor}
\definecolor{linkcolor}{rgb}{0.0,0.3,0.5}
\definecolor{dodgerblue}{HTML}{1E90FF}
\usepackage[unicode, colorlinks=true, linkcolor=linkcolor, citecolor=linkcolor, filecolor=linkcolor,urlcolor=linkcolor, pdfusetitle]{hyperref}
\usepackage[all]{hypcap}
\usepackage[T1]{fontenc}
\usepackage[utf8]{inputenc}
\usepackage{orcidlink}

\usepackage{tensor}
\usepackage{soul}
\usepackage{braket}

\renewcommand{\vec}[1]{\mathbf{#1}}

\makeatletter
\newcommand*{\balancecolsandclearpage}{\close@column@grid \cleardoublepage \twocolumngrid}
\makeatother

\newcommand{\milan}{\affiliation{Dipartimento di Fisica ``G. Occhialini'', Universit\'a degli Studi di Milano-Bicocca, Piazza della Scienza 3, 20126 Milano, Italy}}
\newcommand{\infn}{\affiliation{INFN, Sezione di Milano-Bicocca, Piazza della Scienza 3, 20126 Milano, Italy}}

\begin{document}

\title{Black-hole ringdown with templates capturing spin precession:\\ A reanalysis of GW190521}

\author{Chiara Anselmo$\,$\orcidlink{0009-0008-9328-1177}}
\email{c.anselmo@campus.unimib.it}

\milan \infn

\author{Costantino Pacilio$\,$\orcidlink{0000-0002-8140-4992}}

\milan \infn
\affiliation{Institute of Astrophysics, FORTH, GR-71110, Heraklion, Greece}
\affiliation{INFN, Sezione di Roma, Piazzale Aldo Moro 2, 00185, Roma, Italy}

\author{Davide Gerosa$\,$\orcidlink{0000-0002-0933-3579}}

\milan \infn


\date{\today}

\renewcommand{\arraystretch}{1.5}

\begin{abstract}

The ringdown stage of a binary black-hole merger provides a clean probe of strong-field gravity, as it can be modeled with minimal assumptions.
The quasi--normal-mode frequencies encode the mass and spin of the Kerr black hole remnant, while the mode excitation depends on the progenitor binary.
In this paper, we implement a recently developed amplitude model that captures spin precession in a simulation-based inference pipeline that specifically targets ringdown signals. We present a reanalysis of GW190521---a short-duration, merger-dominated event with conflicting interpretations.
Spin-aligned and precessing analyses at two ringdown start times show that precession induces modest but systematic shifts in inferred parameters and subdominant mode amplitudes, although such ringdown-only analyses provide no strong evidence for precession.
Our results demonstrate the feasibility of physics-informed precessing ringdown modeling, paving the way for the identification of spin precession in gravitational-wave events using solely their ringdown stages, where waveform systematics are expected to be substantially less prominent.

\end{abstract}

\maketitle

\section{Introduction}

The ringdown is the final stage of a binary black-hole (BH) coalescence, during which the merged remnant emits gravitational waves (GWs) as it settles into a stationary state. In general relativity (GR), this equilibrium configuration is uniquely described by a Kerr BH~\cite{1963PhRvL..11..237K,2015CQGra..32l4006T}. Small perturbations of the Kerr spacetime are described by Teukolsky’s equations~\cite{1972PhRvL..29.1114T,1973ApJ...185..635T}, whose solutions correspond to the BH’s quasinormal modes (QNMs)---exponentially damped sinusoids characterized by a discrete complex frequency spectrum~\cite{2009CQGra..26p3001B,1999LRR.....2....2K,1970Natur.227..936V,1971ApJ...170L.105P,1973ApJ...185..649P}. More generally, the full linear response of a perturbed BH is composed of three contributions: a prompt response determined by the initial data, a superposition of QNMs, and a late-time power-law tail arising from backscattering off the spacetime curvature~\cite{2025arXiv250523895B}. In the context of BH mergers, the QNM frequencies depend solely on the remnant’s mass and spin, making them a powerful probe of GR through BH spectroscopy~\cite{1980ApJ...239..292D,2004CQGra..21..787D,2025arXiv250523895B,2018GReGr..50...49B}.

Since the first GW detection~\cite{2016PhRvL.116f1102A}, the LIGO/Virgo/KAGRA (LVK) network has observed about two hundred BH coalescences~\cite{2019PhRvX...9c1040A,2021PhRvX..11b1053A,2024PhRvD.109b2001A,2023PhRvX..13d1039A,2025arXiv250818082T}. Accurate characterization of the ringdown is especially important for events dominated by the merger and ringdown stages, such as GW190521~\cite{2020PhRvL.125j1102A} and GW231123~\cite{2025ApJ...993L..25A}, where the postmerger signal enables a robust study of the source properties. Ringdown-focused analyses are also motivated by waveform systematics: small discrepancies between waveform models can mimic apparent deviations from GR~\cite{2021iSci...24j2577M,2024arXiv240502197G}, and analyses based on different models have sometimes produced inconsistent results~\cite{2022PhRvD.106d4042H,2023CQGra..40i5008B,2025PhRvD.112h4004F,2025ApJ...993L..25A,2025arXiv251102691S,2023PhRvL.131v1402C,2023PhRvD.108f4008S}. Restricting attention to the ringdown, which can be described analytically and depends on fewer modeling assumptions, provides a complementary and often cleaner approach.

While the QNM frequencies are determined entirely by the remnant’s mass and spin, the excitation of the modes, namely their amplitudes and phases, depends on the progenitor binary configuration~\cite{2012PhRvL.109n1102K}. Phenomenological fits to numerical-relativity (NR) simulations map binary parameters, such as mass ratio and spins, to QNM amplitudes, effectively linking premerger dynamics to postmerger observables. Simple models employ post-Newtonian--inspired formulas or results from the extreme-mass-ratio limit, while more sophisticated approaches, including Gaussian process regression (GPR) surrogate models, capture complex dependencies (e.g., spin precession and eccentricity) and provide uncertainty estimates. Several NR-calibrated fits are now available for spin-aligned binaries~\cite{2012PhRvL.109n1102K,2014PhRvD..90l4032L,2018PhRvD..97d4048B,London:2018gaq,2020CQGra..37f5006B,2023PhRvL.130b1001F,2024PhRvD.109d4069C,2024JCAP...10..061C}, and recent GPR models predict QNM amplitudes and phases in that regime~\cite{2024PhRvD.110j3037P,2025PhRvD.112b4077M}. The first NR-calibrated approximant for precessing systems was introduced shortly after by some of us~\cite{2025PhRvD.112d4058N}.

We report the first integration of precessing ringdown templates---constructed from the fits of Ref.~\cite{2025PhRvD.112d4058N}---into a GW parameter estimation framework, and demonstrate their application to GW190521. This event, characterized by a short inspiral and a merger-ringdown--dominated signal, has been extensively studied in the context of overtones, higher harmonics, eccentricity, and precession, both within full inspiral-merger-ringdown (IMR)~\cite{2020PhRvL.125j1102A,2022ApJ...924...79E,2020ApJ...903L...5R,2023NatAs...7...11G,2022NatAs...6..344G,2020ApJ...900L..13A} and ringdown-only~\cite{2023PhRvD.108f4008S,2023PhRvL.131v1402C,2023PhRvD.108f4008S,2020ApJ...900L..13A} frameworks. Our model includes higher harmonics and precession effects, enabling a consistent and physically motivated comparison with previous analyses.

Parameter estimation is performed with simulation-based inference (SBI), also known as likelihood-free Bayesian inference~\cite{2020PNAS..11730055C, 2025arXiv250812939D}. Its application to GW data analysis has been demonstrated both in the frequency~\cite{2023PhRvD.108d4029C} and in the time~\cite{2024PhRvD.110h3010P} domain. In particular, our SBI framework extends the methodology of Ref.~\cite{2024PhRvD.110h3010P}, developed for time-domain ringdown analyses, to precessing templates with physical amplitude models.

We compare a spin-aligned and a precessing ringdown model, assess their performance for two ringdown start times, and quantify the evidence for precession using Bayesian model comparison. Incorporating precession leads to modest but systematic shifts in the inferred binary parameters and QNM amplitudes. The ringdown alone does not provide strong evidence in favor of precession in GW190521. Overall, our findings demonstrate the viability of physics-informed precessing ringdown models and establish a foundation for future high signal-to-noise ratio (SNR) studies of BH spectroscopy.

The remainder of this paper is organized as follows. Section~\ref{sec:ringdown_model} introduces the ringdown model. Section~\ref{sec:methods} describes the analysis methods. Results are presented in Sec.~\ref{sec:results} and discussed in Sec.~\ref{sec:discussion}. We conclude in Sec.~\ref{sec:conclusions}.

\section{Ringdown model}
\label{sec:ringdown_model}

\subsection{Waveform definition and symmetries}
We model the ringdown waveform as a superposition of damped sinusoids corresponding to the QNMs of the remnant BH. This approach isolates the QNM component of the linear response, which is expected to dominate the signal at times sufficiently after the strain peak. The prompt response and the late-time tail are therefore not included, as they are not expected to be statistically significant at currently accessible SNRs in this regime~\cite{1970Natur.227..936V,1978ApJ...224..643C,2025PhRvL.135q1401D,2025PhRvD.112b4003M}.

Each $(\ell,m,n)$ mode contributes a strain component $h_{\ell mn}$, given by
\begin{equation}
\label{eq:h_lmn}
\begin{aligned}
h_{\ell mn} = \dfrac{M}{d_L} & A_{\ell mn} e^{-i(2\pi f_{\ell mn}t + \phi_{\ell mn})} e^{-t/\tau_{\ell mn}} \\
 			& \times \ _{-2}S_{\ell mn}(\iota, \varphi,M_f\chi_f\omega_{\ell mn}) \ .
\end{aligned}
\end{equation}
Here, $M=m_1+m_2$ is the redshifted total mass of the initial binary BH system, $d_L$ is the luminosity distance, and $A_{\ell mn}$ are the amplitudes of excitation of each mode.
The frequencies $f_{\ell mn}$ and damping times $\tau_{\ell mn}$ correspond to the real and imaginary part of the complex mode frequencies, respectively:
\begin{equation}
\omega_{\ell mn} = \omega_R - i\omega_I = 2\pi f_{\ell mn} - \frac{i}{\tau_{\ell mn}} \ .
\end{equation}
These are solutions of Teukolsky's equations~\cite{2015CQGra..32l4006T} and only depend on the mass $M_f$ and spin $\chi_f$ of the remnant.
Finally, $_{-2}S_{\ell mn}(\iota, \varphi,M_f\chi_f\omega_{\ell mn})$ denotes the spin-weighted spheroidal harmonic of that mode, where $\iota$ and $\varphi$ are the inclination and azimuthal angles.

The total waveform is obtained by summing over all relevant modes,
\begin{equation}
\label{eq:h_sum}
\tilde{h} = h_{+}-ih_{\times} = \sum \limits_{\ell mn} h_{\ell mn} \ ,
\end{equation}
where $h_{+}$ and $h_{\times}$ denote the plus and cross GW polarizations, respectively.
For each $(\ell,m,n)$, Teukolsky's equations admit two sets of solutions, the prograde and retrograde modes~\cite{2024PhRvD.109d4069C}. In the above expressions and throughout this work, we neglect the contribution from the retrograde modes. We discuss the motivations and limits of this choice in Sec.~\ref{sec:discussion}.

The measured strain $h$ is the projection of each of the two GW polarizations $h_{+,\times}$ onto the detector through the antenna pattern functions $F_{+,\times}$:
\begin{equation}
\label{eq:s}
h = F_+ h_+ +  F_{\times} h_{\times}\ .
\end{equation}
The pattern functions depend on the sky location $(\alpha,\delta)$ and on the polarization angle $\psi$. They also vary in time, within the timescale of the Earth's rotation; since, for stellar-mass BHs, the typical ringdown signal duration is much shorter ($\sim 0.1 \ s$), we treat $F_{+,\times}$ as constant.

We further simplify the waveform in two ways. First, we replace the spin-weighted spheroidal harmonics $_{-2}S_{\ell mn}$ with spin-weighted spherical harmonics $_{-2}Y_{\ell m}$, thus neglecting spherical-spheroidal mode mixing~\cite{2014PhRvD..90f4012B}. Since we restrict ourselves to the set of modes $\{(2,\pm 2,0),(3,\pm 3,0),(2,\pm 1,0)\}$, which are only mildly affected by mode mixing~\cite{2025PhRvD.112d4058N}, we do not expect our analysis to be significantly biased.

Second, we observe that some parameters are degenerate and can be set to zero without loss of generality.
Since the angular dependence of the spherical harmonics factorizes as a phase,
\begin{equation}
\label{eq:Ylmn}
_{-2}Y_{\ell m}(\iota, \varphi) \equiv \ _{-2}Y_{\ell m}(\iota)e^{im\varphi}~,
\end{equation}
we can reabsorb $\varphi$ in a redefinition of the mode phases, $\phi_{\ell m n}\to\phi_{\ell m n} -m\varphi$.
Similarly, a change in the polarization angle $\psi$ corresponds to a rotation of $h$ by $2\psi$~\cite{2009LRR....12....2S,2023CQGra..40t3001I}. Therefore, we will also reabsorb $\psi$ in a redefinition of the mode phases.
Mind that these degeneracies hold as long as $\phi_{\ell mn}$ are treated as free parameters: if $\phi_{\ell mn}$ were instead modeled as functions of the binary parameters, $(\varphi,\psi)$ would become free parameters of the model.

Finally, only in the case of spin-aligned models, we exploit the following reflection symmetry between $+|m|$ and $-|m|$ modes:
\begin{equation}
\label{eq:spin-aligned_sym}
A_{\ell -|m|n} = (-1)^{\ell} A_{\ell +|m|n} \ .
\end{equation}

\subsection{Amplitude and remnant fits}
We express the QNMs as functions of the binary parameters. We employ two sets of fits, for the mode amplitudes and for the remnant mass and spin, respectively.

\subsubsection{Mode amplitudes}
\label{sec:amps}
For the QNM amplitudes, we adopt the seven-dimensional model \texttt{Prec7dq10\_20M}, first presented in Ref.~\cite{2025PhRvD.112d4058N} by some of the authors and implemented in the \textsc{postmerger} python package~\cite{postmerger}. This model was obtained by training a GPR~\cite{2006gpml.book.....R} on a subset of precessing, quasicircular NR simulations from the 2019 Simulating eXtreme Spacetimes (SXS) catalog~\cite{2019CQGra..36s5006B}. GPR is a nonparametric Bayesian method that represents data as a distribution over functions, providing both mean predictions and associated uncertainties.

The model parameter space is defined by $\{\delta, \boldsymbol{\chi}_1,\boldsymbol{\chi}_2\}$, where
\begin{equation}
\label{eq:delta_from_q}
\delta = \dfrac{q-1}{q+1} \ , \quad q = \dfrac{m_1}{m_2} \geq 1
\end{equation}
and the spins are defined in Cartesian coordinates at the innermost stable circular orbit (ISCO). The fit is calibrated up to mass ratio $q=10$.\footnote{Only a few simulations exist at $q=10$, so we consider the calibration robust up to $q=8$.}
In the spin-aligned limit, the parameter space reduces to three dimensions, $\{\delta,\chi_{1z},\chi_{2z}\}$, by setting the remaining spin components to zero.

The initial binary parameters are provided to the model in the \textit{physical frame}, whose $z$ axis is aligned with the orbital angular momentum. Instead, the resulting amplitudes are expressed in a coordinate system whose $z$ axis is aligned with the spin direction of the remnant BH, called the \textit{ringdown frame}~\cite{2021PhRvD.103h4048F}. This distinction is relevant only in the precessing case and is handled as described in Sec.~\ref{sec:wigner}.

Fitted amplitudes are reported at the reference time
\begin{equation}
t_{\rm start} = 20M + t_{\rm EMOP} \ ,
\end{equation}
where $t_{\rm EMOP}$ is the time of maximum energy-maximized orthogonal projection (EMOP)~\cite{2007PhRvD..76f4034B} of the simulation, extended here to all $(2,m)$ modes (see Appendix~B of Ref.~\cite{2025PhRvD.112d4058N}). For inference purposes, it is instead convenient to define the ringdown starting time relative to the waveform peak, $t_{\rm peak}$, and shift the amplitudes accordingly. The model provides both an estimate of ${\Delta t_{\rm EMOP}=t_{\rm EMOP}-t_{\rm peak}}$ and the possibility to shift the amplitudes with respect to $t_{\rm EMOP}$. Our procedure is described in Sec.~\ref{sec:priors}.

\subsubsection{Final mass and spin}
\label{sec:mass-spin}
To enforce consistency with the amplitude fits, we train a GPR model for final masses and spins in terms of the binary parameters at the ISCO frequency. We use the same training set of NR simulations as in Ref.~\cite{2025PhRvD.112d4058N}.
This implementation has not been presented elsewhere; details and performances are provided in Appendix~\ref{app:mass_spin_fit}.

In particular, going to the ringdown frame only requires knowing the final spin magnitude ${\chi_f = |\boldsymbol{\chi}_f|}$ and the tilt angle
\begin{equation}
\label{eq:theta_f}
\theta_f = \arccos \left( \dfrac{\chi_{fz}}{\chi_f} \right) \ ,
\end{equation}
where $\chi_{fz}$ is the spin component parallel to the orbital angular momentum.
Therefore, we do not need to provide a surrogate for all the three components of $\boldsymbol{\chi}_f$.
We find it convenient to fit the components $\chi_{fz}$ and $\chi_{f\perp}$, where $\chi_{f\perp}$ is the projection of the final spin in the orbital plane, defined by
\begin{equation}
\label{eq:chi_perp}
\chi_f = \sqrt{ \left(\chi_{fz}\right)^2 + \left(\chi_{f\perp}\right)^2 } \ .
\end{equation}
This choice guarantees that both quantities remain within physical ranges and avoids unphysical extrapolations. An initial attempt to fit $(\chi_f,\chi_{fz})$ separately resulted in occasional unphysical extrapolations with $\chi_{fz} > \chi_f$. Fitting $(\chi_{fz}, \chi_{f\perp})$ removes this issue by construction.

\subsection{Wigner rotation of spherical harmonics}
\label{sec:wigner}
Since the mode amplitudes are expressed in the ringdown frame, we write the total strain as~\cite{2025PhRvD.111f4052Z}
\begin{equation}
\tilde h = \sum_{\ell mn} h_{\ell mn}^{\rm (RD)} \, Y_{\ell m}^{\rm (RD)}(\iota',\varphi') \ ,
\end{equation}
where the spherical harmonics in the ringdown frame are related to those in the physical frame through a Wigner rotation,
\begin{equation}
\label{eq:wigner_rot_Y}
Y_{\ell m}^{\rm (RD)}(\iota',\varphi') = \sum_{\mu} D_{m \mu}^{\ell}(\alpha,\beta,\gamma)\; Y_{\ell \mu}^{\rm (phys)}(\iota,\varphi) \ .
\end{equation}
Here, $(\alpha,\beta,\gamma)$ are the Euler angles parametrizing the rotation of the coordinate axes, with $\alpha,\gamma \in [0,2\pi]$ and $\beta \in [0,\pi]$. Following \citeauthor{1960AmJPh..28..408W}~\cite{1960AmJPh..28..408W}, the component rotations are performed in the sequence: a rotation through $\gamma$ about the $z$ axis, then through $\beta$ about the $y$ axis, and finally through $\alpha$ about the $z$ axis.\footnote{Alternative conventions, such as active rotations or different rotation orderings, lead to different sign conventions; see, e.g., Chapter~4 of Ref.~\cite{1988qtam.book.....V} for a comprehensive discussion. Our choice follows the passive Wigner convention of Ref.~\cite{1960AmJPh..28..408W}, which fixes the signs used below.}

The corresponding Wigner D matrix is given by
\begin{equation}
\label{eq:wigner_d-matrix}
D_{m \mu}^{\ell}(\alpha,\beta,\gamma) = e^{im\gamma}\, d_{m \mu}^{\ell}(\beta)\, e^{i\mu \alpha} \ ,
\end{equation}
where $d_{m \mu}^{\ell}(\beta)$ is the Wigner's (small) $d$ matrix, see Eq.~(15.27) of~Ref.~\cite{1960AmJPh..28..408W},
\begin{align}
&d_{m \mu}^{\ell}(\beta)= \left[(\ell + m)!(\ell -m)!(\ell + \mu)!(\ell -\mu)! \right]^{\frac{1}{2}} \notag  \\
&\times \sum \limits_{s=s_{\rm min}}^{s_{\rm max}} \left[ \dfrac{(-1)^s\left(\cos \tfrac{\beta}{2}\right)^{2\ell+\mu-m-2s}\left(\sin \tfrac{\beta}{2}\right)^{m-\mu+2s}}{(\ell+\mu-s)!s!(m-\mu+s)!(\ell-m-s)!} \right] \,,
\end{align}
with summation limits ${s_{\rm min} = \max (0,\mu-m)}$ and ${s_{\rm max} = \min (\ell+\mu,\ell-m)}$. Note that our definition of $d_{m \mu}^{\ell}(\beta)$ maps into $d_{m \mu}^{\ell}(-\beta)\equiv d_{\mu m}^{\ell}(\beta)$ when following the alternative conventions of Refs.~\cite{2011PhRvD..84b4046S,2021PhRvD.103h4048F}.

The geometric interpretation of the Euler angles is as follows: the second angle, $\beta$, corresponds to the inclination $\theta_f$ between the $z$ axis in the physical frame, parallel to the orbital angular momentum, and the final spin vector $\boldsymbol{\chi_f}$; the first angle, $\gamma$, can be reabsorbed into the QNM phases $\phi_{\ell mn}$; the third angle, $\alpha$, is degenerate with the azimuthal angle $\varphi$ [see Eq.~\eqref{eq:Ylmn}]. Moreover, the condition that $\varphi'$ is reabsorbed in a redefinition of the QNM phases implies $\varphi=-\gamma$ in Eq.~\eqref{eq:wigner_rot_Y}. As a result, we can simplify the rotation matrices so that only ${\beta=\theta_f}$ appears explicitly,
\begin{equation}
\tilde h = \sum_{\ell mn} \sum_{\mu} h_{\ell mn}^{\rm (RD)} \, d_{m \mu}^{\ell}(\theta_f) \, Y_{\ell \mu}^{\rm (phys)}(\iota,0) \ .
\end{equation}

\section{Methods}
\label{sec:methods}

\subsection{TSNPE}
In addition to being an established Bayesian inference framework, SBI offers practical advantages for the present analysis compared to standard likelihood-based sampling approaches. In the current implementation of the \textsc{postmerger} package, waveform evaluations are sufficiently expensive ($t_{\rm eval}\sim 1~\rm s$) that repeated likelihood estimations within a traditional Markov chain Monte Carlo (MCMC) sampler would lead to a substantial computational cost at the scale required for this study. By contrast, SBI allows waveform simulations to be generated in large batches, amortizing the cost of forward-model evaluations and making the inference computationally tractable.

Among the broad class of SBI algorithms, we adopt truncated sequential neural posterior estimation (TSNPE)~\cite{2022arXiv221004815D} as our inference method. In \emph{neural posterior estimation}, parameters are sampled from the prior $\pi(\theta)$ to generate simulated data $x(\theta)$, and a neural density estimator $q_{\phi}(\theta | x)$ with learnable parameters $\phi$ is trained to directly approximate the posterior. \emph{Sequential} neural posterior estimation~\cite{2016arXiv160506376P,2019arXiv190507488G} specializes this to the case of a particular observation $x_{\rm obs}$ and proceeds in multiple adaptive rounds: in the first round, training samples are drawn from the prior $\pi(\theta)$; in subsequent rounds, they are drawn from a proposal distribution $\tilde{\pi}(\theta)$ that is updated to be denser in regions of high posterior density $q_{\phi}(\theta | x_{\rm obs})$, the latter approximated from the previous round.
In particular, TSNPE updates the prior with a sequential truncation scheme: at each round, the original training set is augmented with new simulations, drawn from the high-density region of the approximate posterior---see Refs.~\cite{2022arXiv221004815D,2024PhRvD.110j3037P} for further details.

Our implementation of TSNPE closely follows that of Ref.~\cite{2024PhRvD.110h3010P}, which uses the \textsc{PyTorch}-based~\cite{2019arXiv191201703P} \textsc{sbi} package~\cite{2025JOSS...10.7754B}. The neural density estimator is modeled as a neural spline flow~\cite{2019arXiv191202762P,2019arXiv190604032D,2020arXiv200803312G}. For this paper, we have made minor updates to the neural-network architecture as detailed Appendix~\ref{sec:nn}.

\subsection{Setup}
When analyzing ringdown, one needs to truncate the signal in order to isolate the postmerger part of the waveform. This truncation requires fixing the ringdown starting time. Since different sky locations correspond to different delays in the arrival time across the detector network, it is also necessary to fix the source sky location, ensuring that the same stretch of data is analyzed throughout the inference~\cite{2021arXiv210705609I}.\footnote{While this issue commonly arises in time-domain analyses---where truncation can be implemented without spectral leakage---it would similarly affect frequency-domain analyses involving truncated signals. An alternative frequency-domain ringdown analysis that alleviates these difficulties has been proposed by \citeauthor{2021PhRvD.104l3034F}~\cite{2021PhRvD.104l3034F}.}
We adopt from Ref.~\cite{2024EPJC...84..233G} the waveform peak time $t_{\rm peak}$ at the Hanford detector (expressed in the GPS time scale) and the sky location of GW190521:
\begin{equation}
\begin{aligned}
t_{\rm peak} &= 1242442967.4287114 \ , \\
\alpha &= 0.16416072898840461~\rm rad \ , \\
\delta &= -1.1434284511661967~\rm rad \ .
\end{aligned}
\end{equation}
These values were obtained from the IMR LVK analysis for GW190521 using the posterior samples corresponding to the NRSur7dq4 model. In particular, the starting time corresponds to the time that maximizes ${h_+^2 + h_{\times}^2}$, while the sky position is estimated as the median of the posterior samples.

The time delays at the other detectors are computed using the \texttt{detector} module from \textsc{PyCBC}~\cite{pycbc}.

We analyze the ringdown of GW190521 starting at 6~ms and 12~ms after $t_{\rm peak}$. For a binary with redshifted mass $\sim 250~M_\odot$, these times correspond to $\sim 5M$ and $\sim 10M$, respectively.
While a starting time of 12~ms is more conservative with respect to the physical model, 6~ms is less conservative but has a higher SNR.
Both times constitute an extrapolation with respect to the nominal calibration time of the amplitude fits, which are reported at $20M$ after $t_{\rm EMOP}$. This extrapolation serves to amplify the SNR and is not expected to lead to appreciable biases at low-to-moderate SNRs~\cite{2018PhRvD..98j4020C,2021PhRvD.103l2002A,2025PhRvD.112h4080A}.
We find consistent recoveries of the parameters (total mass, mass ratio, spins, and luminosity distance) at both starting times. We interpret this consistency as an \textit{a posteriori} internal validation of the robustness of our analysis.
Moreover, at the SNRs considered here (see Sec.~\ref{sec:snr}), we expect the waveform template to be robust for mismatches ${\mathcal{M} \lesssim 0.1}$~\cite{2024arXiv240106845T}, which is at least an order of magnitude higher than the typical mismatches reported in Ref.~\cite{2025PhRvD.112d4058N}. However, note that this assumes that there are no systematics between the template and the true waveforms; the role of systematics is discussed in Sec.~\ref{sec:discussion}.

The duration of the analysis segment after each start time is set to 0.1~s for all detectors, as in Ref.~\cite{2020ApJ...900L..13A}.
We load the strain data using the \texttt{catalog} module from \textsc{PyCBC} and downsample them to $2048~\rm Hz$ with the \texttt{decimate} method from \texttt{scipy.signal}. See Ref.~\cite{2025PhRvD.111d4070S} for recent developments on data conditioning techniques, and the Supplement to Ref.~\cite{2025PhRvL.135k1403A} (Appendix IIB2) for small differences that can arise when analyzing different signal lengths. We whiten the time-domain strain following the procedure described in Ref.~\cite{2024PhRvD.110h3010P} (see their Appendix~A1).

Our ringdown template includes the six prograde fundamental modes
\begin{equation}
{(\ell,m,n)= \{(2,\pm2,0),(3,\pm3,0),(2,\pm1,0)\}} \ .
\end{equation}
The initial BH spins $\boldsymbol{\chi_i}$ are expressed in spherical coordinates, $\{\chi_i,\gamma_i,\phi_i\}$, where $\chi_i=|\boldsymbol{\chi_i}|$ is the spin magnitude and $\gamma_i$ and $\phi_i$ denote the polar and azimuthal angles, respectively. The QNM frequencies, assuming GR, are determined from $(M_f,\chi_f)$ using the \textsc{pykerr} package~\cite{pykerr}. Altogether, the model depends on $16$ parameters:
\begin{equation}
\theta = \left\{d_L,\cos \iota,M,q,\chi_i,\cos \gamma_i,\phi_i, \phi_{\ell mn} \right\} \ ,
\end{equation}
where $i=1,2$ and $(\ell,m,n)$ run over the included modes.
In the spin-aligned limit, the spin vectors $\boldsymbol{\chi}_i$ reduce to their $z$ components, $\chi_{iz}$, lowering the number of dimensions to $12$.

\subsection{Priors}
\label{sec:priors}
We infer the posterior distribution using priors that are uniform in the parameters and ranges listed in Table~\ref{tab:prior_ranges}. In particular, in the precessing model, we adopt priors that are uniform in the spin magnitudes $\chi_i$ and isotropic in orientation, as commonly done in GW astronomy; this corresponds to uniform priors in $\cos \gamma_i$. In the spin-aligned model, we instead use a uniform prior on the aligned spin components $\chi_{iz}$.
Note that, in the precessing model, uniform priors in $\chi_{i}$ and $\cos\gamma_i$ induce highly nonuniform priors on $\chi_{iz}$. Therefore, to ensure consistency when comparing results with the posterior samples from the spin-aligned model, we reweight the latter so that their implied prior on $\chi_{iz}$ matches that of the isotropic case, as described in Appendix~\ref{app:rew_spins}.

For the masses, we adopt uniform priors in total mass $M$ and mass ratio $q$, chosen for sampling efficiency. To consistently compare our results with IMR analyses, which assume priors uniform in the component masses $(m_1, m_2)$, we reweight our posterior samples following the procedure detailed in Appendix~\ref{app:rew_masses}.

\begin{table}
\centering
\begin{tabular}{l@{\hspace{16em}}r}
\hline\hline
Parameter & Prior range \\
\hline
\multicolumn{2}{c}{Shared} \\
$d_L^3 \ (\rm Gpc^3)$ & $[10^{-3},10^3]$ \\
$\cos \iota$ & $[-1,1]$ \\
$M \ (M_{\odot})$ & $[100,500]$ \\
$q$ & $[1,10]$ \\
$\phi_{\ell mn}$ & $[0,2\pi]$ \\

\multicolumn{2}{c}{Precessing model} \\

$\chi_i$ & $[0,0.99]$ \\
$\cos \gamma_i$ & $[-1,1]$ \\
$\phi_i$ & $[0,2\pi]$ \\

\multicolumn{2}{c}{Spin-aligned model} \\

$\chi_{iz}$ & $[-0.99,0.99]$ \\
\hline\hline
\end{tabular}
\caption{Prior ranges adopted for the parameter estimation. All parameters are assigned uniform priors over the intervals listed. The \textit{Shared} block (top) lists parameters common to both the spin-aligned and precessing models, while the treatment of the spins differs between the two cases (center and bottom).}
\label{tab:prior_ranges}
\end{table}

Importantly at inference time, for each proposed total mass $M$, the amplitude fits are consistently evaluated at a time
\begin{equation}
\label{eq:t_eval}
t_{\rm eval}=t_{\rm peak}+\frac{t_{\rm start}}{M}=t_{\rm EMOP}-\Delta t_{\rm EMOP}+\frac{t_{\rm start}}{M}
\end{equation}
when expressed in NR simulation units. Since the amplitude fits are calibrated at ${t_{\rm EMOP}+20}$, this amounts to rescale the amplitudes by a factor ${\exp{t_{\rm start}/M-20-\Delta t_{\rm EMOP}}}$. In this way, the start time adapts consistently to the sampled value of $M$, while the physical onset of the ringdown remains identical across all realizations, ensuring internal consistency in the parameter estimation.

\section{Results}
\label{sec:results}

We begin by presenting the posterior distributions for the binary parameters inferred with the spin-aligned and precessing ringdown models, followed by the corresponding QNM amplitudes derived from these posteriors.
For completeness, we also report the posteriors for the remnant mass and spin in Appendix~\ref{app:remnant}, including results from this work and from other analyses.
Throughout this Section, and unless otherwise specified, binary properties and QNM amplitudes are computed using the mean predictions of the GPR fits of Ref.~\cite{2025PhRvD.112d4058N}; the impact of including the GPR variances is examined separately in Appendix~\ref{app:gpr_sigmas}. Finally, we compare the two models by computing their Bayes factor and evaluate the ringdown SNRs associated with the chosen start times. Modeling assumptions and potential limitations relevant to the interpretation of these results are discussed in Sec.~\ref{sec:discussion}.

\subsection{Binary parameters}
\begin{figure*}
\centering
\includegraphics[width=\textwidth]{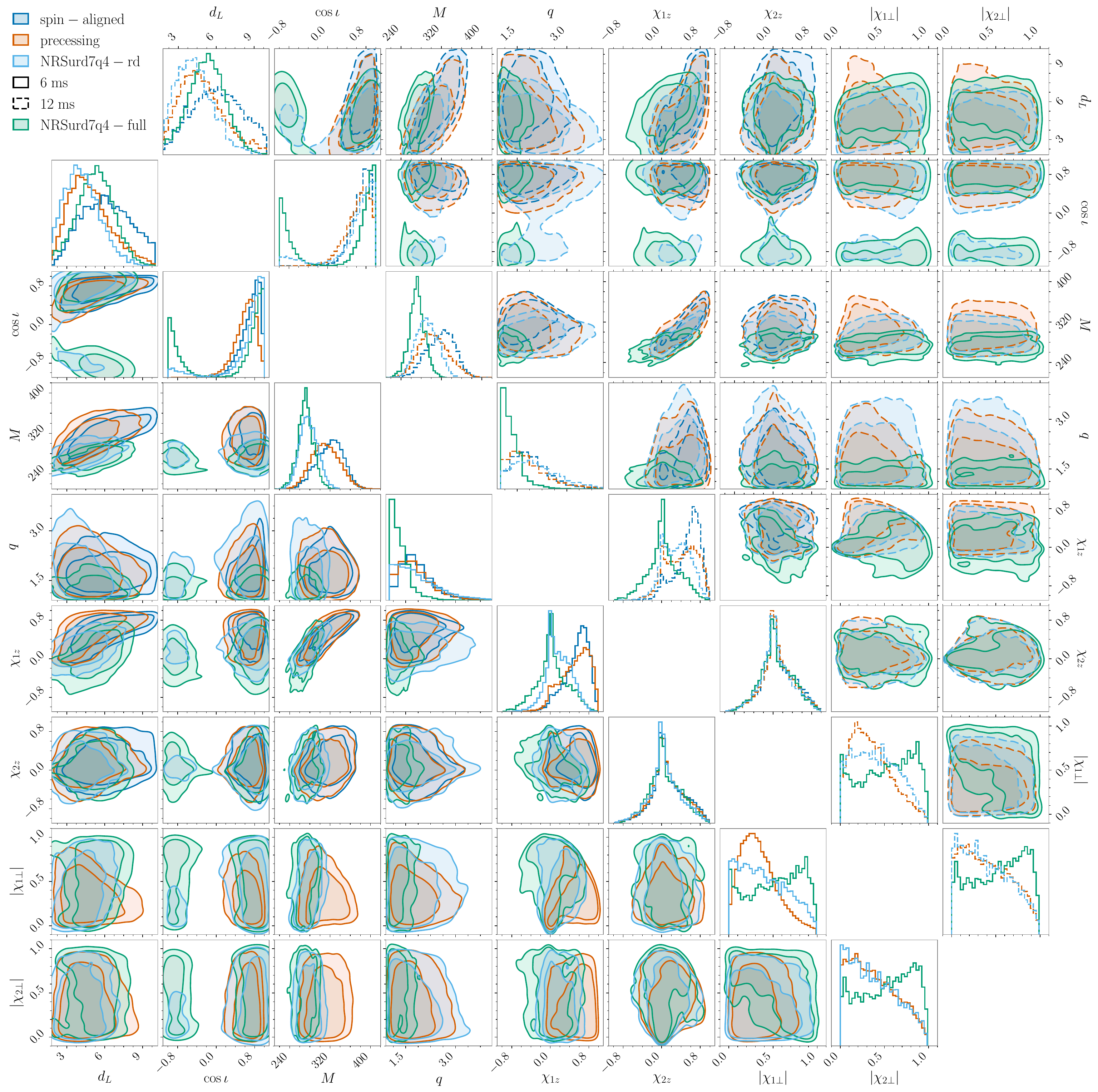}
\caption{Joint posterior distributions for the binary parameters inferred with the spin-aligned (blue) and precessing (orange) ringdown models, using the GPR mean predictions for the remnant mass and spin and for the QNM amplitudes. The lower (upper) triangular panels correspond to a ringdown start time of 6~ms (12~ms) after the strain peak, shown with solid (dashed) lines.
The IMR posterior from the LVK analysis~\cite{2020PhRvL.125j1102A,2020ApJ...900L..13A} (green), and the results from Ref.~\cite{2024PhRvD.109b4024M} (light blue), both obtained with the \texttt{NRSur7dq4} model~\cite{2019PhRvR...1c3015V}, are overlaid for comparison.
All contours represent the 68\% and 90\% credible regions.}
\label{fig:corner_1_pred}
\end{figure*}

Our parameter estimation yields a $16$- ($12$-) dimensional posterior distribution in the precessing (spin-aligned) case.
Figure~\ref{fig:corner_1_pred} shows the joint posterior distributions for ringdown start times of 6~ms (lower triangle) and 12~ms (upper triangle), over the following parameters:
\begin{enumerate}
\item $\{d_L,\cos\iota,M,q\}$, common to both models;
\item $\{\chi_{1z},\chi_{2z}\}$, which are reweighted spin components in the spin-aligned case, and computed as ${\chi_{iz} = \chi_i \cos\gamma_i}$ in the precessing case;
\item $\{|\chi_{1\perp}|,|\chi_{2\perp}|\}$, present only in the precessing model, and obtained from $(\chi_i,\chi_{iz})$ as ${|\chi_{i\perp}|=\sqrt{(\chi_i)^2-(\chi_{iz})^2}}$.
\end{enumerate}

We compare our results with the IMR parameter estimation reported by the LVK Collaboration~\cite{2020PhRvL.125j1102A,2020ApJ...900L..13A}, which makes use of the full signal and includes precession effects. We use the IMR samples produced with the \textsc{NRSur7dq4} model~\cite{2019PhRvR...1c3015V}, which was the preferred waveform approximant in the GW190521 LVK analyses~\cite{2020ApJ...900L..13A,2020PhRvL.125j1102A}.
We also include the results of Ref.~\cite{2024PhRvD.109b4024M}, where the data from the LVK detector network were truncated at different times from the peak and analyzed in the time domain using the \texttt{NRSur7dq4} model. Posterior samples are taken from the dataset at Ref.~\cite{miller_2023_8349282}, using start times of $5M_{\rm ref}$ ($10M_{\rm ref}$) to match our 6~ms (12~ms) analysis ($M_{\rm ref}\simeq 1.27~\rm ms$, see Sec.~II of Ref.~\cite{2024PhRvD.109b4024M} for details). Since those samples were obtained with uniform priors on total mass $M$, mass ratio $q$, and luminosity distance $d_L$, they have been reweighted following the procedure described in Appendix~\ref{app:rew}.

Across all analyses and start times, we find consistent trends in several key parameters. 
The inclination $\iota$ is similarly constrained in all models, although the IMR posterior for $\cos\iota$ is mildly bimodal. This feature does not appear in our runs, as we fix the sky location $(\alpha,\delta)$ to values compatible with $\cos\iota \simeq 1$ in the IMR run, breaking the degeneracy.
A similar behavior is observed in the $\iota$ posterior of Ref.~\cite{2024PhRvD.109b4024M}, where the sky location is also fixed.
The mass ratio $q$ favors relatively symmetric binaries ($q \lesssim 2$), and the secondary spin $\chi_{2z}$ remains largely unconstrained, reflecting the isotropic prior.\footnote{In comparing spins, we note that the IMR values are defined at ${t=-\infty}$, while ours correspond to ISCO. The resulting differences are subdominant compared to statistical uncertainties at current SNRs~\cite{2022PhRvD.105b4076M}.}

The most notable differences between the spin-aligned and precessing analyses arise in the luminosity distance $d_L$, total mass $M$, and primary aligned spin $\chi_{1z}$. Including precession systematically shifts $d_L$ to smaller values---smaller even than the IMR result.
The $d_L$ posterior from Ref.~\cite{2024PhRvD.109b4024M} is in good agreement with our precessing results at both start times.
The shift in $M$ follows the same trend, though in this case the precessing posterior moves closer to the IMR estimate.
For this parameter, the results of Ref.~\cite{2024PhRvD.109b4024M} are consistent with the IMR analysis at 6~ms, but show a similar shift at 12~ms.
The effect on $\chi_{1z}$ becomes significant in the 12~ms analysis, where precession reduces the inferred value relative to the spin-aligned model.
Similarly, while consistent with the IMR analysis at 6~ms, the posterior from Ref.~\cite{2024PhRvD.109b4024M} shows better agreement with our precessing results at 12~ms.
Finally, the in-plane spin components $(|\chi_{1\perp}|,|\chi_{2\perp}|)$ remain largely unconstrained, with the posterior favoring very small values more strongly than the IMR analysis.
This is consistent with the results of Ref.~\cite{2024PhRvD.109b4024M} at both start times. The discrepancy between our analysis and those performed with \texttt{NRSur7dq4} at 6~ms may be related to the absence of the $(2,2,1)$ mode in our model~\cite{2023PhRvD.108j4020B,2026PhRvL.136d1403A}. More generally, we observe that, as the start time moves further away from the merger, the agreement between the two ringdown analyses (ours and that of Ref.~\cite{2024PhRvD.109b4024M}) improves.

\subsection{Mode amplitudes}
\begin{figure*}[hbtp]
\centering
\includegraphics[width=0.7\textwidth]{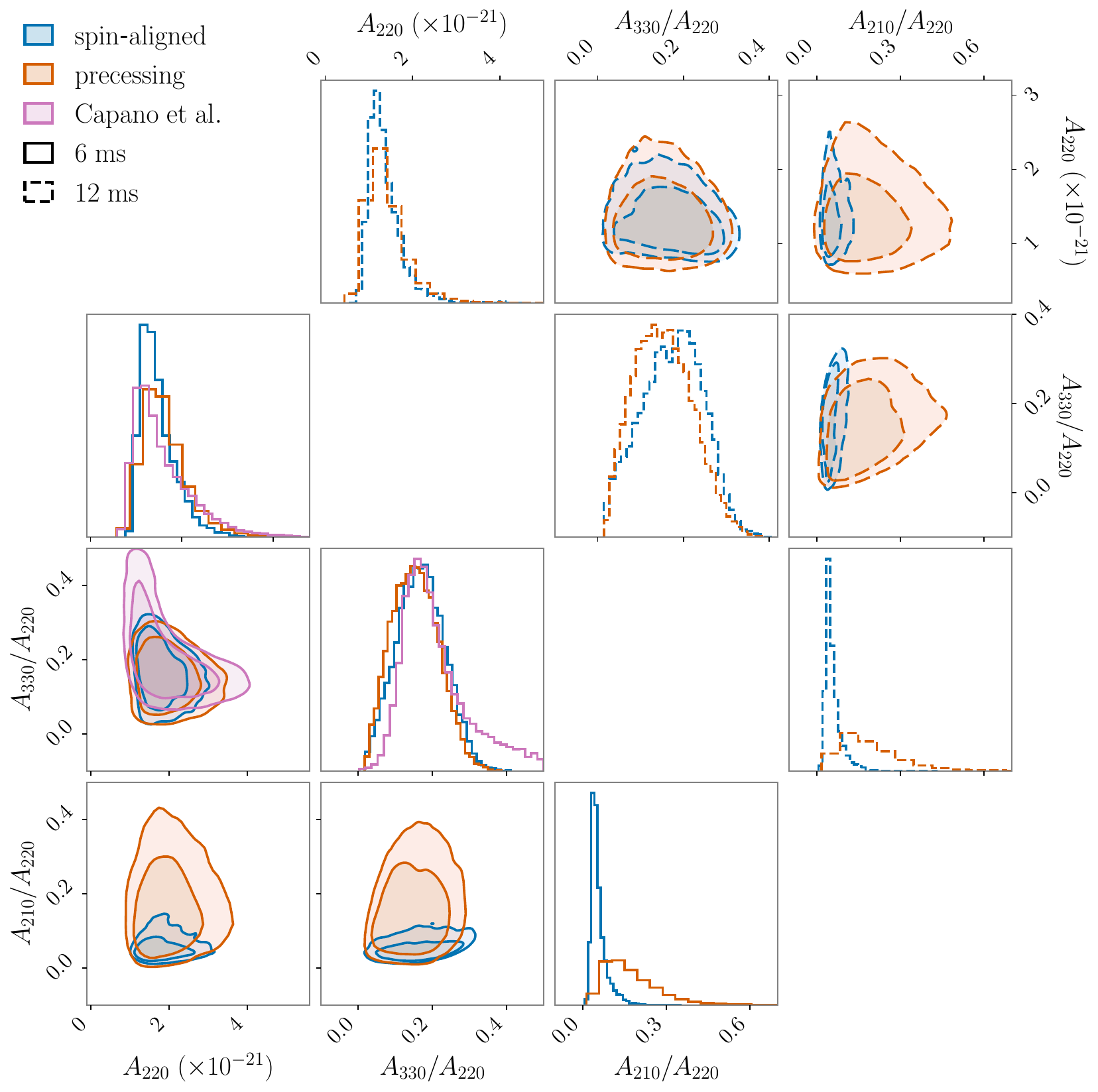}
\caption{Joint posterior distributions for the QNM amplitudes inferred from the spin-aligned (blue) and precessing (orange) ringdown models, using the GPR mean predictions for the remnant mass and spin and for the QNM amplitudes. We show the absolute amplitude of the dominant $(2,2,0)$ mode, $A_{220}$, together with the relative amplitudes $A_{330}/A_{220}$ and $A_{210}/A_{220}$. The lower (upper) triangular panels correspond to a ringdown start time of 6~ms (12~ms) after the strain peak, shown with solid (dashed) lines. For the 6~ms case, we overlay the results of Ref.~\cite{2023PhRvL.131v1402C} (pink), obtained using a $(2,2,0)+(3,3,0)$ Kerr model. Contours indicate the 68\% and 90\% credible regions.}
\label{fig:corner_2_pred}
\end{figure*}

Starting from the posterior samples, we compute the QNM amplitudes $A_{\ell +|m|n}$ using the same phenomenological fits as in the main analysis. Figure~\ref{fig:corner_2_pred} shows the resulting distributions of $A_{220}$ as well as the relative amplitudes $A_{330}/A_{220}$ and $A_{210}/A_{220}$. The lower (upper) triangle corresponds to the 6~ms (12~ms) case. For the 6~ms analysis, we also include the results of Ref.~\cite{2023PhRvL.131v1402C}, obtained with a $(2,2,0)+(3,3,0)$ Kerr model and publicly available at Ref.~\cite{collin_capano_2023_10056546}.

The spin-aligned posteriors for $A_{220}$ and $A_{210}/A_{220}$ are narrower compared to the case where precession is included, reflecting the smaller parameter space. Both models yield comparable results for $A_{220}$, but the precessing posteriors favor lower $A_{330}$ and higher $A_{210}$, indicating a modest enhancement of subdominant mode excitation due to precession.

The spin-aligned relative amplitude $A_{330}/A_{220}$ agrees with the results of Ref.~\cite{2023PhRvL.131v1402C}. While that analysis is agnostic in the amplitudes, our results provide an independent, physics-informed indication of the relevance of the $(3,3,0)$ mode in the ringdown of GW190521. Finally, for both start times, the $(2,2,0)$ mode remains dominant ($A_{210}<A_{220}$). This differs from the highly precessional interpretation of Ref.~\cite{2023PhRvD.108f4008S}, which reported support for $A_{210}>A_{220}$ at early start times. In the time window considered here, our analysis favors the standard mode hierarchy, while still indicating that precession can leave measurable imprints on subdominant QNM amplitudes.

A direct comparison with other ringdown analyses in the literature is not always straightforward, due to differences in modeling choices and time windows.
In particular, Ref.~\cite{2023PhRvD.108f4008S} does not provide posteriors obtained with templates that simultaneously include both the $(3,3,0)$ and $(2,1,0)$ modes, while Ref.~\cite{2024EPJC...84..233G} finds the strongest support for the $(3,3,0)$ mode at premerger start times. These differences limit the extent to which quantitative comparisons can be drawn.
The comparison with Ref.~\cite{2023PhRvL.131v1402C} is more direct. Although our template includes the $(2,1,0)$ mode, the posterior of the corresponding reconstructed amplitude has non-negligible support at zero, effectively reducing its impact to the overall inference. In this sense, the two analyses are expected to be broadly consistent in the reconstruction of the $(3,3,0)$ mode amplitude.

\subsection{Bayes factors}
To quantify the evidence for precession, we compute the Bayes factor $\mathcal{B}_{\rm P}^{\rm SA}$ between the spin-aligned model $M_{\rm SA}$ and the precessing model $M_{\rm P}$. The two models are nested,
\begin{equation}
M_{\rm SA}\equiv M_{\rm P} \wedge \left(\cos^2\gamma_1=1,\cos^2\gamma_2=1\right) \ .
\end{equation}
Therefore, we can evaluate the Bayes factor through the Savage-Dickey ratio
\begin{equation}
    \label{eq:sd:1}
    \mathcal{B}_{\rm P}^{\rm SA}=\sum_{a,b=\pm1}\frac{p(\cos\gamma_1=a,\cos\gamma_2=b|x_{\rm obs},M_{\rm P})}{\pi(\cos\gamma_1=a,\cos\gamma_2=b|M_{\rm P})}\frac{1}{4|a||b|} \ .
\end{equation}

Since we sample isotropically in the component spins, the prior density at the denominator is exactly $1/4$. To evaluate the posterior density at the numerator, we use a two-dimensional kernel density estimator (KDE) on the posterior samples for $\left(\cos\gamma_1,\cos\gamma_2\right)$, with Gaussian kernels and a bandwidth of $0.05$. To prevent density leakage, we reflect the samples about the upper boundaries $\cos\gamma_i=1$ and about the upper corner ${\cos\gamma_1=\cos\gamma_2=1}$ of the domain, where the two-dimensional posterior peaks. We then apply a correction factor of $4$ to the KDE, to compensate for the broadening of the domain.

We find  ${\log_{10}\mathcal{B}_{\rm P}^{\rm SA}\approx0.28}$ at ${t_{\rm start}=t_{\rm peak}+{\rm 12~ms}}$, and ${\log_{10}\mathcal{B}_{\rm P}^{\rm SA}\approx0.63}$ at ${t_{\rm start}=t_{\rm peak}+{\rm 6~ms}}$. These numbers correspond to weak evidence against precession according to Jeffreys's scale~\cite{jeffreys1998theory}. This is consistent with previous results by \citeauthor{2024PhRvD.109b4024M}~\cite{2024PhRvD.109b4024M}, which show that the evidence for precession found in IMR analyses~\cite{2020PhRvL.125j1102A,2020ApJ...900L..13A} mainly comes from the last inspiral cycles before merger (see also Ref.~\cite{2021PhRvD.104j3018B}).

\subsection{Signal-to-noise ratios}
\label{sec:snr}
Let us recall the noise-weighted inner product between two time series $x$ and $y$,
\begin{equation}
\braket{x|y}=\sum_{i=1}^N (x_{\rm white})_i(y_{\rm white})_i  \ ,
\end{equation}
where $N$ is the number of data points, and $x_{\rm white}$ denotes the original time series $x$ after whitening it with the noise power spectral density---and similarly for $y_{\rm white}$. Whitening is performed directly in the time domain, see Refs.~\cite{2021arXiv210705609I,2024PhRvD.110h3010P}.
We report the matched-filter SNR~\cite{2021arXiv210705609I,2024EPJC...84..233G} 
\begin{equation}
    \rho_{\rm mf}=\frac{\braket{x_{\rm obs}|h(\boldsymbol{\theta})}}{\sqrt{\braket{h(\boldsymbol{\theta})|h(\boldsymbol{\theta})}}} \ .
\end{equation}

For the spin-aligned model, we find ${\rho_{\rm mf}=10.7^{+0.2}_{-0.4}}$ at ${t_{\rm start}=t_{\rm peak}+{\rm 6~ms}}$ and ${\rho_{\rm mf}={9.1}_{-0.2}^{+0.5}}$ at ${t_{\rm start}=t_{\rm peak}+{\rm 12~ms}}$, where the lower and upper limits denote 90\% confidence interval over $10^4$ posterior samples. For the precessing model, we find ${\rho_{\rm mf}=10.4^{+0.3}_{-0.9}}$ at ${t_{\rm start}=t_{\rm peak}+{\rm 6~ms}}$ and ${\rho_{\rm mf}={8.7}_{-1.0}^{+0.4}}$ at ${t_{\rm start}=t_{\rm peak}+{\rm 12~ms}}$. The low signal strength at 12~ms past $t_{\rm peak}$ motivated us to extrapolate the model at times earlier than its nominal calibration time $t_{\rm EMOP}+20M$.

\section{Discussion}
\label{sec:discussion}

In this Section, we outline some caveats that should be kept in mind when interpreting the results presented above.

\paragraph*{Modeling assumptions.}
The amplitude model adopted in this work~\cite{2025PhRvD.112d4058N} does not yet include all potentially relevant physical ingredients.

First, retrograde modes can contribute non-negligibly to the ringdown signal for certain configurations of mass ratio and spin orientations~\cite{2025PhRvD.111f4052Z,2023PhRvD.107j4035H,2021PhRvD.103j4048D}. Their inclusion is part of planned future developments. For the event considered here, neglecting retrograde modes is expected to be a reasonable approximation if the final spin tilt angle is small. Using posterior samples and our surrogate model (Sec.~\ref{sec:mass-spin}), we find that the tilt angle satisfies $\theta_f \lesssim \pi/8$ at 90\% credible probability, supporting this approximation for the present analysis.

Second, the current implementation does not include the $(3,2,0)$ mode. Reference~\cite{2023PhRvD.108f4008S} found this mode to be consistent with the NRSur7dq4 posteriors for GW190521. While we recover the dominant qualitative features of the ringdown without this mode, its omission limits the extent to which our framework can reproduce analyses that explicitly include it. Incorporating the $(3,2,0)$ mode---and, more generally, extending the QNM content---is a natural direction for future improvements of the model.

\paragraph*{Reliability of SBI.}
Simulation-based inference introduces potential sources of uncertainty beyond standard sampling error, including biases arising from the neural posterior approximation and training limitations~\cite{2021arXiv211006581H,2021arXiv210104653L}. In the present work, modeling uncertainties dominate over sampling-related uncertainties. Once the amplitude model is improved, a more thorough study of SBI-related uncertainties will be needed; we leave this for future work.

\paragraph*{GPR uncertainties.}
The main results in Sec.~\ref{sec:results} are obtained using the mean predictions of the GPR fits. A fully consistent treatment would exploit the full GPR predictive distribution. As discussed in Appendix~\ref{app:gpr_sigmas}, sampling from the GPR posterior currently leads to very broad uncertainties in some derived quantities. For this reason, we present the mean-based results as our primary findings and defer a more complete treatment of GPR-induced uncertainties to future work.

\section{Conclusions}
\label{sec:conclusions}

In this work, we presented the first implementation of ringdown templates that incorporate spin precession within a GW parameter estimation framework, and applied them to GW190521. Using physics-informed fits for the remnant and the QNM excitation, we carried out spin-aligned and precessing ringdown analyses of GW190521 at two start times (6 and 12~ms after the strain peak).

Including precession leads to consistent shifts toward smaller luminosity distances, slightly smaller total masses, and reduced primary aligned spins, with the effect most pronounced in the 12~ms analysis. The in-plane spin components remain weakly constrained, as expected at the available SNR. The corresponding QNM amplitude posteriors show that both models yield compatible estimates for the dominant $(2,2,0)$ amplitude, while precession modestly enhances subdominant excitation, lowering $A_{330}/A_{220}$ and increasing $A_{210}/A_{220}$. In all cases, the $(2,2,0)$ mode remains dominant, consistent with standard expectations.
Bayesian model selection yields $\log_{10}\mathcal{B}_{\rm P}^{\rm SA}\simeq0.63$ (6~ms) and $\simeq0.28$ (12~ms), indicating that ringdown data are not informative enough to draw firm conclusions about precession.

While full IMR analyses consider the entirety of the signal and are therefore more informative, the benefit of a ringdown-only investigation is that waveform systematics are expected to be much less severe. Provided one is sufficiently away from the peak of the strain (and how far is the subject of active investigation~\cite{2018PhRvD..97j4065B,2023PhRvD.108j4020B,2025PhRvD.112h4076V,2025arXiv251102915C}), perturbation theory unambiguously prescribes the strain reported in Eq.~\eqref{eq:h_lmn}. On the other hand, IMR analyses for high-mass, merger-ringdown--dominated signals are subject to strong systematics. The most emblematic case is GW231123~\cite{2025ApJ...993L..25A}, whose parameters present order-unity differences when analyzed with different state-of-the-art waveform models.

Although the ringdown waveform is prescribed by perturbation theory, our analysis still inherits potential systematics from the amplitude fits, thus shifting the signpost toward the need for increased accuracy in those surrogate models. To this end, the phenomenological fits for the ringdown parameters can be retrained using the recently released SXS catalog~\cite{2025CQGra..42s5017S}, as well as public results from other groups. Future versions of the model could also include additional QNMs beyond the fundamental $\{(2,\pm 2,0),(3,\pm 3,0),(2,\pm 1,0)\}$, or incorporate corrections for orbital eccentricity~\cite{2024JCAP...10..061C}. The latter effect is often neglected under the assumption that BH binaries circularize efficiently through GW emission before merger, but even such residual eccentricity can introduce systematics in the interpretation of GW signals~\cite{2024PhRvD.110f3012F}.
\citeauthor{2024JCAP...10..061C}~\cite{2024JCAP...10..061C} showed that noncircular orbits can substantially affect QNM excitation, with relative changes in mode amplitudes of up to 50\%.
A more systematic and uniform comparison with existing ringdown analyses, accounting for differences in waveform content and analysis choices, will be important to further assess the robustness and consistency of current interpretations.

Most recently, \citeauthor{2025arXiv251013954D}~\cite{2025arXiv251013954D,2025arXiv251011783D} presented Bayesian extractions of ringdown modes from NR simulations, using a likelihood that encodes information on the numerical resolution of the simulations themselves. Propagating such uncertainties upstream to analyses like ours would effectively turn potentially undetected systematic errors into statistical ones.

\acknowledgements

We thank Swetha Baghwat, Francesco Nobili, Gregorio Carullo, and Riccardo Buscicchio for discussions.
C.A., C.P., and D.G. are supported by 
ERC Starting Grant No.~945155--GWmining, 
Cariplo Foundation Grant No.~2021-0555, 
MUR PRIN Grant No.~2022-Z9X4XS, 
Italian-French University (UIF/UFI) Grant No.~2025-C3-386,
MUR Grant ``Progetto Dipartimenti di Eccellenza 2023-2027'' (BiCoQ),
the ICSC National Research Centre funded by NextGenerationEU,
and the INFN TEONGRAV initiative.
C.P. is supported by ERC Starting Grant No.~101117624--MMMonsters, MUR FIS2 Advanced Grant ET-NOW (CUP:~B53C25001080001).
D.G. is supported by MSCA Fellowship No.~101149270--ProtoBH and
MUR Young Researchers Grant No. SOE2024-0000125.
Computational work was performed 
at CINECA with allocations through INFN and the University of Milano-Bicocca, 
and at NVIDIA with allocations through the Academic Grant program.

\section*{Data availability}

The ringdown amplitude model is available at Ref.~\cite{postmerger}.
Posterior samples for the inference runs presented in this paper are available at Ref.~\cite{datarelease}.

\appendix

\begin{figure*}
    \centering
  \includegraphics[width=0.8\linewidth]{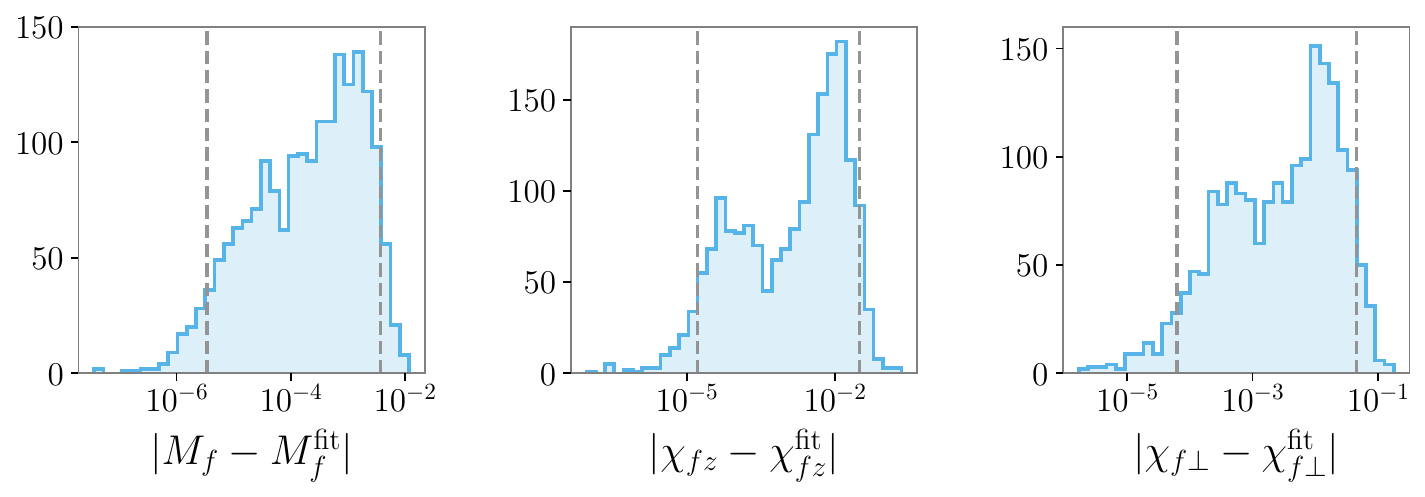}
    \caption{Residual errors on our remnant GPR fits for $\left(M_f,\chi_{fz},\chi_{f\perp}\right)$, respectively, computed via 10-fold cross validation. Dashed vertical lines delimit the 90\% density regions.}
    \label{fig:remnant:residuals}
\end{figure*}

\begin{table}
\centering
\begin{tabular}{l@{\hspace{4em}}c@{\hspace{4em}}c}
\hline\hline
Quantity & RMSE (mean) & RMSE (std.~dev.) \\
\hline
$M_f$              & $0.002$ & $0.002$ \\
$\chi_{fz}$     & $0.014$ & $0.003$ \\
$\delta\chi_{f\perp}$ & $0.018$ & $0.004$ \\
\hline\hline
\end{tabular}
\caption{Root-mean-squared errors (RMSEs) for the fitted quantities ${(M_f, \chi_{fz}, \delta\chi_{f\perp})}$, reported as mean and standard deviation.}
\label{tab:rms_errors}
\end{table}

\section{Performance of the mass-spin fit}
\label{app:mass_spin_fit}

We describe the construction of our surrogate model for predicting the final mass and spin of the remnant BH, as a function of the mass ratio and the component spins at ISCO. Unlike the ringdown amplitude model, these fits for the remnant properties have not been presented elsewhere. The input parameters are ${\{\delta,\chi_{+z},\chi_{-z},\chi_{1x},\chi_{1y},\chi_{2x},\chi_{2y}\}}$, where
\begin{equation}
\delta=\dfrac{(q-1)}{(q+1)} \ , \quad \chi_{\pm z}=\dfrac{q\chi_{1z}\pm\chi_{2z}}{1+q}
\end{equation}
and we order the component masses such that ${q=m_1/m_2 \geq 1}$.

To predict the final mass $M_f$, we build a GPR for the quantity $1-M_f$. To predict the magnitude $\chi_f$ of the final spin, we first build a GPR for the component $\chi_{fz}$ of the final spin along the $z$ direction in the physical frame. Then, we build a second GPR for the quantity $\delta\chi_\perp=\chi_{f\perp}-\chi_\perp$, where
\begin{equation}
    \chi_{f\perp}=\sqrt{\left(\chi_f\right)^2-\left(\chi_{fz}\right)^2}
\end{equation}
is the magnitude of the final spin on the orbital plane,
\begin{equation}
    \chi_\perp=\frac{\sqrt{||\vec{S}_1+\vec{S}_2||^2-\left(S_{1z}+S_{2z}\right)^2}}{(m_1+m_2)^2}
\end{equation}
is the total spin momentum on the orbital plane at ISCO, normalized by the component masses, and ${\vec{S}_i=m_i^2\vec{\boldsymbol{\chi}}_i}$ is the physical spin of the $i$th progenitor. Then, the magnitude of the final spin is reconstructed as
\begin{equation}
\chi_f=\sqrt{\left(\chi_{fz}\right)^2+\left(\chi_{f\perp}\right)^2} \ .
\end{equation}
This construction ensures that the geometric condition $\chi_f\geq\chi_{fz}$ is always satisfied.

All surrogates are built by first subtracting a linear fit to the quantity of interest, then training a GPR on the corresponding residuals. Similarly to Ref.~\cite{2025PhRvD.112d4058N}, we use a combination of a radial basis function kernel times a constant kernel, summed to a white kernel.

Table~\ref{tab:rms_errors} shows the RMSEs for the three fitted quantities $(M_f, \chi_{fz},\chi_{f\perp})$, computed via tenfold cross-validation~\cite{2009eds..book.....L}. Figure~\ref{fig:remnant:residuals} shows the corresponding residual errors: to avoid overfitting, the residuals are evaluated at validation time, then the ten validation batches are merged together. Overall, test errors lie well within the measurement uncertainties of our parameter estimates, making the model suitable for this work.

\begin{figure*}[hbtp]
    \centering
  \includegraphics[width=0.8\linewidth]{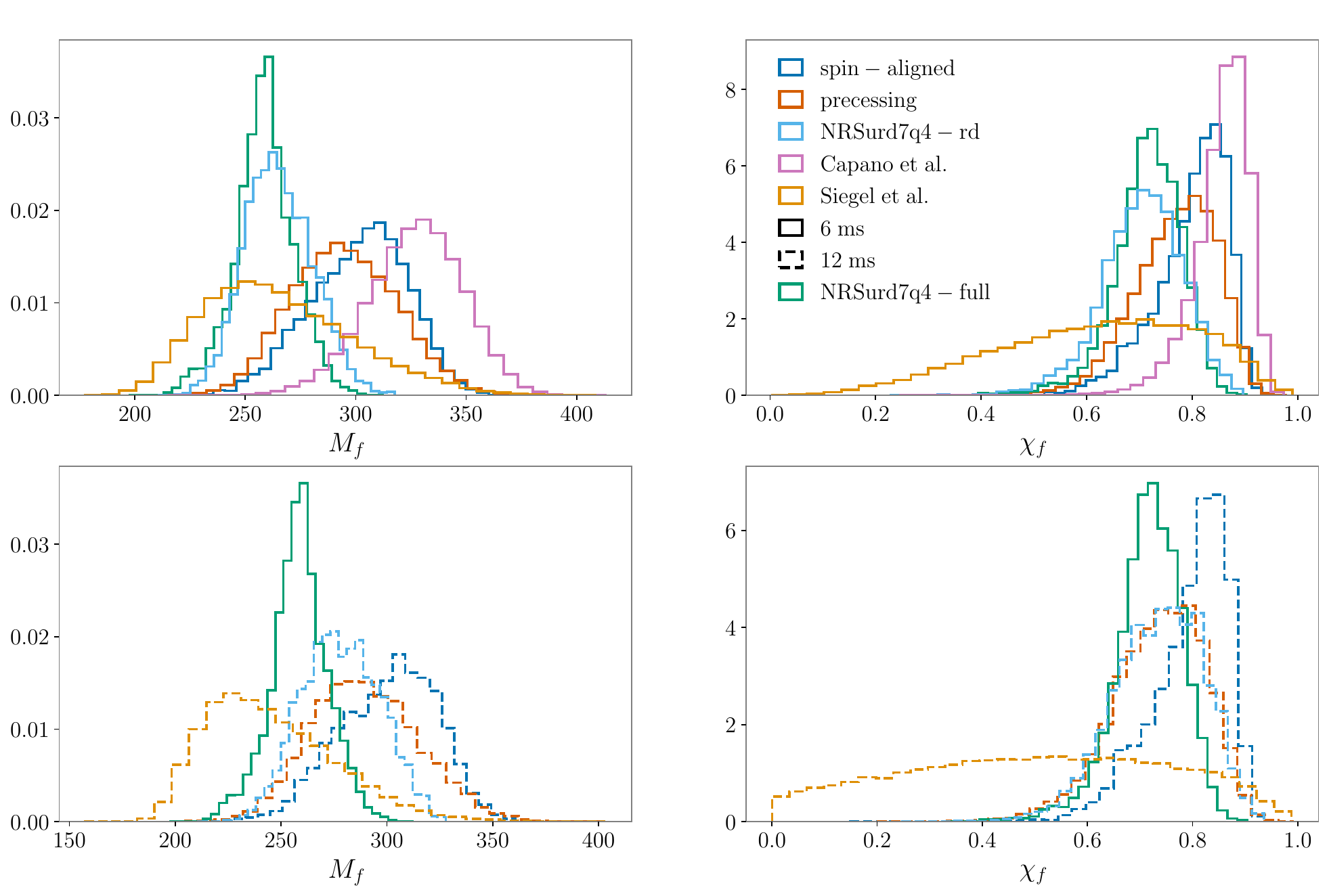}
    \caption{
Posteriors for the remnant mass $M_f$ and spin $\chi_f$, inferred with the spin-aligned (blue) and precessing (orange) ringdown models using the GPR mean predictions.
For comparison, we also show the IMR posterior from the LVK analysis~\cite{2020PhRvL.125j1102A,2020ApJ...900L..13A} (green), and the results of Ref.~\cite{2024PhRvD.109b4024M} (light blue), both obtained with the \texttt{NRSur7dq4} model~\cite{2019PhRvR...1c3015V}.
Additional comparisons include Ref.~\cite{2023PhRvL.131v1402C} (pink), based on a $(2,2,0)+(3,3,0)$ Kerr model, and Ref.~\cite{2023PhRvD.108f4008S} (dark yellow), using a $(2,2,0)+(2,1,0)+(3,2,0)$ template.
The top (bottom) panels correspond to a ringdown start time of 6~ms (12~ms) after the strain peak, shown with solid (dashed) lines.}
    \label{fig:final_mass_spin}
\end{figure*}

\section{Reweighting procedure}
\label{app:rew}

As discussed in Sec.~\ref{sec:priors}, the precessing and spin-aligned models adopt different prior choices for the BH spins, and our sampling of the mass parameters employs a different parametrization from that used in IMR analyses. To enable consistent comparisons between models, we reweight the posterior samples to match the corresponding target priors, following a similar procedure to Appendix~A of \cite{2018arXiv180510457L}. For each posterior sample, the weight is given by the ratio between the target and initial priors,
\begin{equation}
w(x) = \frac{\pi_{\mathrm{target}}(x)}{\pi(x)} \ .
\end{equation}

\subsection{Spins}
\label{app:rew_spins}
In the precessing model, the priors are uniform in spin magnitude $\chi$ and in $\cos\gamma$ (corresponding to isotropic orientations), whereas in the spin-aligned case the prior is uniform in the aligned component $\chi_z$, with the in-plane components set to zero. To consistently compare the two cases, we reweight the spin-aligned posterior samples so that their prior on $\chi_z$ matches the prior induced by isotropy.

We first compute the target prior as a function of $\chi_z = \chi\cos\gamma$,
\begin{equation}
\begin{aligned}
\pi_{\rm target} (\chi_z) = \int_0^1 d\chi \int_{-1}^1 d(\cos\gamma) \, \pi(\chi)\pi(\cos\gamma) \\
\times \ \delta(\chi_z - \chi \cos\gamma) \ .
\end{aligned}
\end{equation}
Since $\chi \sim \mathcal{U}(0,1)$ and $\cos\gamma \sim \mathcal{U}(-1,1)$, we have
\begin{equation}
\pi_{\rm target} (\chi_z) = \frac{1}{2}\int_0^1 d\chi \int_{-1}^1 d(\cos\gamma) \ \delta(\chi_z - \chi \cos\gamma) \ .
\end{equation}
Using the identity $\delta(f(x)) = \delta(x-x_0)/|f'(x_0)|$ with ${x=\cos\gamma}$, we find
\begin{equation}
\delta(\chi_z - \chi \cos\gamma) = \frac{\delta\!\left(\cos\gamma - \frac{\chi_z}{\chi}\right)}{\chi} \ .
\end{equation}
The $\delta$ function restricts the integration to $|\chi_z| \leq \chi$, leading to
\begin{equation}
\pi_{\rm target} (\chi_z) = \frac{1}{2}\int_{|\chi_z|}^1 \frac{d\chi}{\chi}
= -\frac{1}{2}\ln |\chi_z| \ .
\end{equation}
The initial prior is uniform, $\pi(\chi_z) = 1/2$, giving
\begin{equation}
\label{eq:weights_spin}
w(\chi_z) = \frac{\pi_{\rm target}(\chi_z)}{\pi(\chi_z)}
= - \ln |\chi_z| \ .
\end{equation}
For both spins $(\chi_{1z},\chi_{2z})$, the total weight is $w = w_1 w_2$, with each $w_i$ given by Eq.~\eqref{eq:weights_spin}. The resulting weights are then normalized.

\subsection{Masses}
\label{app:rew_masses}
Our sampling employs uniform priors in total mass $M$ and mass ratio $q$ for numerical efficiency, while IMR analyses typically adopt priors uniform in the component masses $(m_1,m_2)$. To reweight our samples, we derive the corresponding Jacobian between $(m_1,m_2)$ and $(M,q)$.

For $q \geq 1$, one has ${m_1=Mq/(1+q)}$ and ${m_2=M/(1+q)}$. The Jacobian determinant of the transformation is thus
\begin{equation}
|\det J|=M/(1+q)^2 \ .
\end{equation}
We write the induced density in $(M,q)$ as
\begin{equation}
\label{eq:pi_target_masses}
\pi_{\rm target}(M,q) \propto \dfrac{M}{(1+q)^2} \ .
\end{equation}

In addition, we must enforce the constraint that the component masses lie within the physical prior bounds, which is tantamount to
\begin{equation}
\label{eq:M_constraint}
m_{\rm low}(1+q) \leq M \leq m_{\rm high}(1+q)/q \ ,
\end{equation}
where $m_{\rm low}$ and $m_{\rm high}$ are the lower and upper bounds on the component masses. In particular, to reweight our initial posteriors, we use $m_{\rm low}=25$ and $m_{\rm high}=250$, consistent with the prior ranges on $M$ and $q$ (see Table~\ref{tab:prior_ranges}). Therefore, the reweighting factors are given by Eq.~\eqref{eq:pi_target_masses}, subject to the constraint of Eq.~\eqref{eq:M_constraint}.

\section{Remnant mass and spin posteriors}
\label{app:remnant}

Figure~\ref{fig:final_mass_spin} shows the posterior distributions for the remnant mass and spin obtained with our spin-aligned and precessing models, for ringdown start times of 6~ms (top panels) and 12~ms (bottom panels).
As in the main text, we compare our results with those of Refs.~\cite{2020PhRvL.125j1102A,2020ApJ...900L..13A,2024PhRvD.109b4024M}, based on the \texttt{NRSur7dq4} model. For the latter, remnant quantities are evaluated consistently using the associated remnant fits~\cite{2019PhRvR...1c3015V}.
In the 6~ms case, we also include the posteriors from Ref.~\cite{2023PhRvL.131v1402C}, obtained with a $(2,2,0)+(3,3,0)$ ringdown model.
Finally, we show the remnant mass and spin reported in Ref.~\cite{2023PhRvD.108f4008S}, obtained with the ${(2,2,0)+(2,1,0)+(3,2,0)}$ template, which provides their best agreement with \texttt{NRSur7dq4}. We use their results corresponding to start times of 6.35~ms and 12.7~ms, approximately matching our 6~ms and 12~ms cases, respectively.

\begin{table}[h!]
	\begin{tabular}{l@{\hspace{12em}}r}
	\hline\hline
	\multicolumn{2}{c}{Embedding network: Fully connected} \\
	\hline
	\texttt{input\_dim} & $612$ \\
	\texttt{output\_dim} & $192$ \\
	\texttt{nun\_hidden\_layers} & $1$ \\
	\texttt{hidden\_dim} & $50$ \\[5pt]
	\hline\hline 
	\multicolumn{2}{c}{Density estimator: Neural spline flow} \\
	\hline
	\texttt{num\_blocks} & $2$ \\
	\texttt{hidden\_features} & $150$ \\
	\texttt{num\_transforms} & $5$ \\
	\texttt{num\_bins} & $10$ \\
	\texttt{batch\_normalization} & True \\[5pt]
	\hline\hline
	\multicolumn{2}{c}{Training hyperparameters} \\
	\hline
	\texttt{num\_simulations} & $300k$ \\
	\texttt{batch\_size} & $512$ \\
	\texttt{batch\_norm} & True \\
	\texttt{learning\_rate} & $10^{-3}$ \\
	\texttt{truncation\_quantile} & $10^{-4}$ \\
	\texttt{stopping\_volume\_ratio} & $0.8$ \\
	\texttt{validation\_fraction} & $0.1$ \\
	\texttt{varying\_noise} & True \\
	\hline\hline
	\end{tabular}
	\vspace{5mm}
\caption{Summary of the neural-network architecture and training hyperparameters used in our SBI framework. Compared to Ref.~\cite{2024PhRvD.110j3037P}, this setup employs a simpler embedding network and a larger training set.}
\label{tab:nn_details}
\end{table}

\begin{figure*}
\centering
\includegraphics[width=\textwidth]{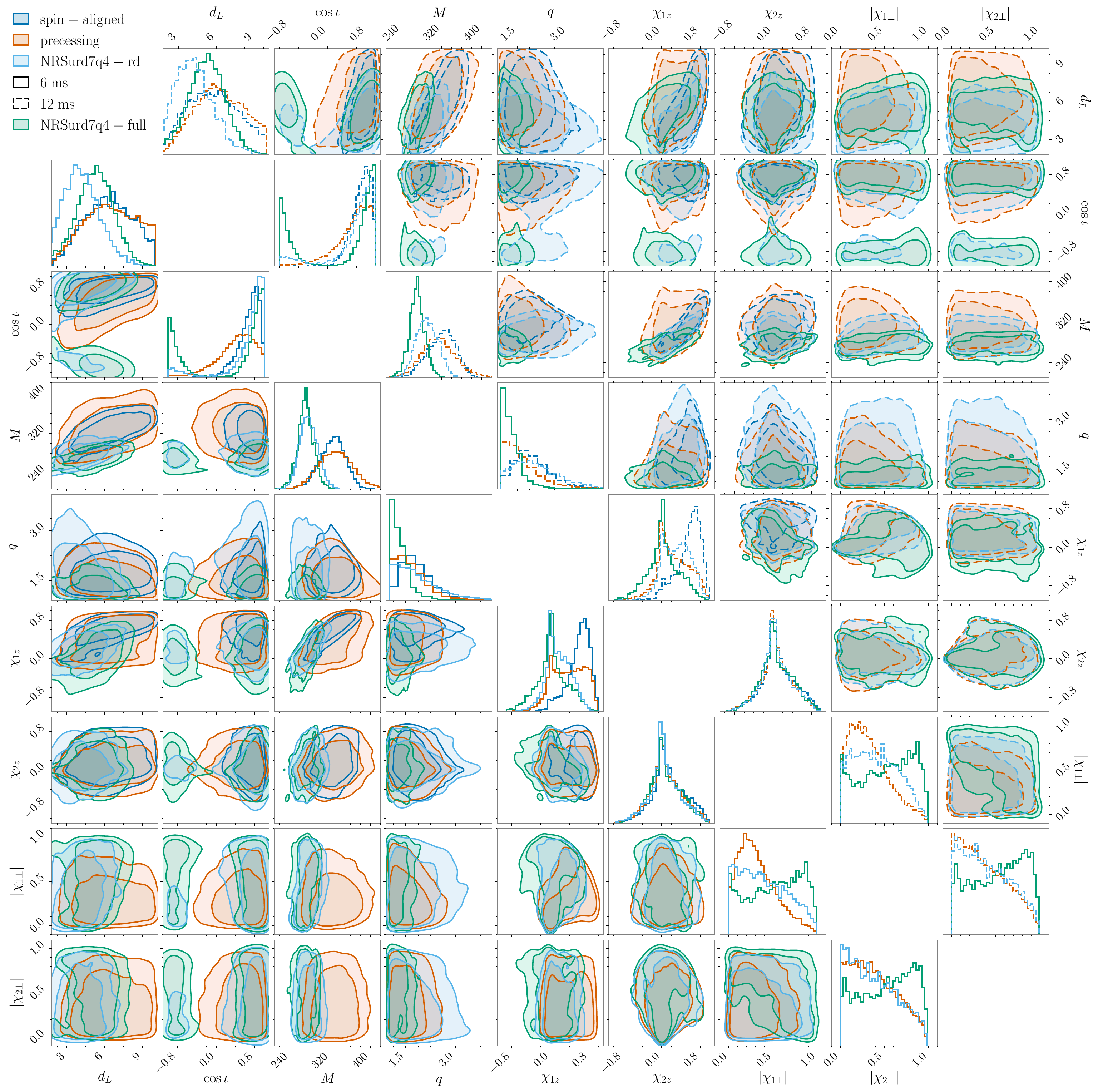}
\caption{Joint posterior distributions for the binary parameters inferred with the spin-aligned (blue) and precessing (orange) ringdown models, using randomly sampled GPR realizations of the remnant mass and spin and of the QNM amplitudes. The lower (upper) triangular panels correspond to a ringdown start time of 6~ms (12~ms) after the strain peak, shown with solid (dashed) lines.
The IMR posterior from the LVK analysis~\cite{2020PhRvL.125j1102A,2020ApJ...900L..13A} (green), and the results from Ref.~\cite{2024PhRvD.109b4024M} (light blue), both obtained with the \texttt{NRSur7dq4} model~\cite{2019PhRvR...1c3015V}, are overlaid for comparison.
All contours represent the 68\% and 90\% credible regions.}
\label{fig:corner_1_sample}
\end{figure*}

\begin{figure*}[hbtp]
\centering
\includegraphics[width=0.7\textwidth]{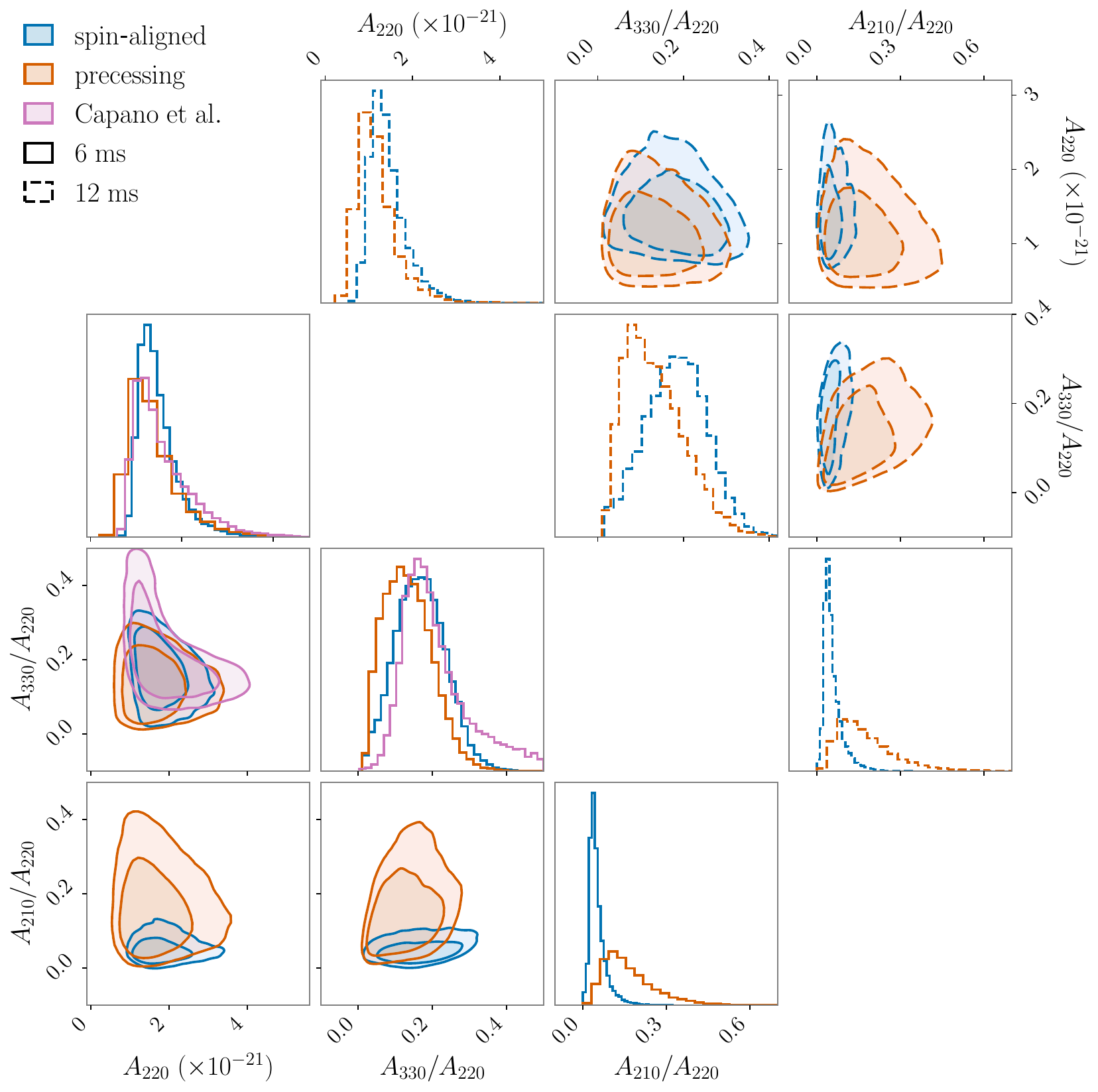}
\caption{Joint posterior distributions for the QNM amplitudes inferred from the spin-aligned (blue) and precessing (orange) ringdown models, using randomly sampled GPR realizations of the remnant mass and spin and of the QNM amplitudes. We show the absolute amplitude of the dominant $(2,2,0)$ mode, $A_{220}$, together with the relative amplitudes $A_{330}/A_{220}$ and $A_{210}/A_{220}$. The lower (upper) triangular panels correspond to a ringdown start time of 6~ms (12~ms) after the strain peak, shown with solid (dashed) lines. For the 6~ms case, we overlay the results of Ref.~\cite{2023PhRvL.131v1402C} (pink), obtained using a $(2,2,0)+(3,3,0)$ Kerr model. Contours indicate the 68\% and 90\% credible regions.}
\label{fig:corner_2_sample}
\end{figure*}

\section{Neural-network architecture and training}
\label{sec:nn}

The hyperparameters of the neural-network architecture and training are summarized in Table~\ref{tab:nn_details}. With respect to Ref.~\cite{2024PhRvD.110j3037P} (see their Table~III), we adopt a simpler embedding network, consisting of a single hidden layer with $50$ neurons (instead of two hidden layers with $150$ neurons each), but with a slightly larger output dimension ($192$ versus $128$). We also train the network using a larger number of simulations ($3\times 10^5$ fixed, compared to at most $1\times 10^5$). These changes resulted in a more robust inference in our case.

\section{Analysis with GPR uncertainties}
\label{app:gpr_sigmas}

In this Appendix, we assess the impact of the GPR uncertainties on the inferred binary parameters and QNM amplitudes by sampling predictions from Gaussian distributions using the predicted means and standard deviations. The rest of the analysis is identical to that of the main text.

Note that our SBI implementation is set to optimize the loss function~\cite{2020arXiv200803312G}
\begin{equation}
    \label{eq:sbi:loss}
    L = \mathbb{E}\left[-\log q_\phi(\theta|x_{\rm obs})\right] \ ,
\end{equation}
where the expectation value is a Monte Carlo sum over $\theta\sim \pi(\theta)$ and $x_{\rm obs}\sim p(d|\theta)$. When we sample the GPR values randomly, we are in fact taking also the expectation value over the latent functional space $\boldsymbol{\lambda}$ of the GPR model. To see this, we rewrite the loss function as
\begin{equation}
    \label{eq:sbi:loss:2}
    L = \mathbb{E}\left[-\int d\boldsymbol{\lambda}~\pi_{\rm GPR}(\boldsymbol{\lambda})~\log q_\phi(\theta,|x_{\rm obs},\boldsymbol{\lambda})\right] \ ,
\end{equation}
and replace the integral over the measure $d\boldsymbol{\lambda}~\pi_{\rm GPR}(\boldsymbol{\lambda})$ with a Monte Carlo integral over random GPR samples. This shows that we are effectively marginalizing the likelihood over the GPR uncertainties. SBI allows us to perform this marginalization by simply extending the scope of the Monte Carlo expectation values in Eq.~\eqref{eq:sbi:loss}. On the contrary, in a typical MCMC sampler, one would need to marginalize the likelihood via a Monte Carlo integral at each step, thus introducing an additional computational cost.

Figure~\ref{fig:corner_1_sample} shows the analog of Fig.~\ref{fig:corner_1_pred}, now obtained using randomly sampled GPR realizations rather than the GPR mean predictions.
As expected, including the GPR uncertainties broadens several posteriors. This effect is stronger for the luminosity distance $d_L$ and total mass $M$, for which the differences between the spin-aligned and precessing models largely disappear. For $\chi_{1z}$, the precessing posterior at 6~ms becomes consistent with the (lower-SNR) 12~ms result in Fig.~\ref{fig:corner_1_pred}. The inclination $\cos\iota$ and mass ratio $q$ also show mild widening, but remain broadly consistent with the mean-prediction results. The secondary spin component $\chi_{2z}$ and the in-plane spins $(|\chi_{1\perp}|,|\chi_{2\perp}|)$ remain essentially unconstrained, so sampling the GPRs produces negligible changes relative to Fig.~\ref{fig:corner_1_pred}.

An analogous comparison can be made for the QNM amplitudes. Figure~\ref{fig:corner_2_sample} shows the analog of Fig.~\ref{fig:corner_2_pred}, now using GPR samples.
Including the GPR uncertainties tends to accentuate the differences between the spin-aligned and precessing models. In particular, the $A_{220}$ posterior becomes less consistent between the two models, and the precessing posterior for $A_{330}/A_{220}$ shifts to lower values even at 6~ms (whereas in Fig.~\ref{fig:corner_2_pred} this occurred only at 12~ms).

Repeating the calculation of the Bayes factor for these GPR-sampled runs, we obtain ${\log_{10}\mathcal{B}_{\rm P}^{\rm SA}\approx0.11}$ at ${t_{\rm start}=t_{\rm peak}+{\rm 12~ms}}$, and ${\log_{10}\mathcal{B}_{\rm P}^{\rm SA}\approx0.41}$ at ${t_{\rm start}=t_{\rm peak}+{\rm 6~ms}}$.
Similarly, the matched-filter SNRs for the spin-aligned model are ${\rho_{\rm mf}=10.7^{+0.2}_{-0.4}}$ at ${t_{\rm start}=t_{\rm peak}+{\rm 6~ms}}$ and ${\rho_{\rm mf}={9.1}_{-0.4}^{+0.2}}$ at ${t_{\rm start}=t_{\rm peak}+{\rm 12~ms}}$. For the precessing model, we find lower SNRs but with much broader uncertainties: ${\rho_{\rm mf}=8.1_{-8.2}^{+2.4}}$ at ${t_{\rm start}=t_{\rm peak}+{\rm 6~ms}}$ and ${\rho_{\rm mf}=7.3_{-5.9}^{+1.6}}$ at ${t_{\rm start}=t_{\rm peak}+{\rm 12~ms}}$.
These large uncertainties arise from sampling the full GPR posterior; using the mean predictions, as in the main analysis, yields more stable SNR estimates.

\bibliography{ringsbi}

\end{document}